\RequirePackage{fix-cm}
\documentclass[smallextended]{svjour3}       
\usepackage{algorithmic}
\usepackage{amsmath}
\usepackage{amssymb}
\smartqed  
\usepackage{graphicx}
\usepackage{algorithm}
\usepackage{booktabs}
\begin{document}

\title{Enhancing the Practical Reliability of Shor’s Quantum Algorithm via Generalized Period Decomposition: Theory and Large-Scale Empirical Validation}

\author{Chih-Chen Liao \and
        Chia-Hsin Liu  \and
        Yun-Cheng Tsai
}


\institute{Chih-Chen Liao and Yun-Cheng Tsai\at
              Department of Technology Application and Human Resource Development \\
              National Taiwan Normal University \\
              No.162, He-Ping East Road Sec1, Taipei 10610, Taiwan \\              
              Yun-Cheng Tsai is the corresponding author. \email{pecu@ntnu.edu.tw}           
           \and
           Chia-Hsin Liu \at
           Department of Mathematics \\
           Taipei Municipal University of Education
}

\maketitle

\begin{abstract}
This work presents a generalized period decomposition approach, significantly improving the practical reliability of Shor’s quantum factoring algorithm. Although Shor’s algorithm theoretically enables polynomial-time integer factorization, its real-world performance heavily depends on stringent conditions related to the period obtained via quantum phase estimation. Our generalized decomposition method relaxes these conditions by systematically exploiting arbitrary divisors of the obtained period, effectively broadening the applicability of each quantum execution. Extensive classical simulations were performed to empirically validate our approach, involving over one million test cases across integers ranging from 2 to 8 digits. The proposed method achieved near-perfect success rates—exceeding 99.998\% for 7-digit numbers and 99.999\% for 8-digit numbers—significantly surpassing traditional and recently improved variants of Shor’s algorithm. Crucially, this improvement is achieved without compromising the algorithm’s polynomial-time complexity and integrates seamlessly with existing quantum computational frameworks. Moreover, our method enhances the efficiency of quantum resource usage by minimizing unnecessary repetitions, making it particularly relevant for quantum cryptanalysis with noisy intermediate-scale quantum (NISQ) devices. This study thus provides both theoretical advancements and substantial practical benefits, contributing meaningfully to the field of quantum algorithm research and the broader field of quantum information processing.
\end{abstract}

\section{Introduction}
Quantum computing represents a revolutionary paradigm, offering exponential speedups for some computational issues that are intractable on classical computers. A pivotal achievement in quantum information science is Shor's algorithm, which factors large integers in polynomial time by exploiting quantum parallelism~\cite{nielsen2010quantum,shor1999polynomial}. This algorithm has profound implications for modern cryptography, especially RSA encryption, whose security depends on the presumed hardness of factoring large composite numbers using classical techniques~\cite{rivest1978method,chen2016report}. As quantum hardware advances toward practical viability, enhancing the reliability of Shor's algorithm is crucial for quantum cryptanalysis and broader applications in quantum information processing.

While the theoretical underpinnings of Shor's algorithm are solid, practical implementations often face challenges stemming from its reliance on specific properties of the period derived via quantum phase estimation~\cite{shor1999polynomial,gidney2021factor}. The algorithm's success hinges on obtaining an even period that meets particular mathematical criteria; failure to do so necessitates restarting with a new random base, incurring significant inefficiencies~\cite{nielsen2010quantum,dong2023improving}. These iterations are especially burdensome in contemporary quantum systems, constrained by short coherence times and high operational costs~\cite{leander2002improving,dalvi2024one}.

To mitigate these issues, recent research has introduced enhancements targeting specific limitations in Shor's algorithm. For example, Dong et al. (2023) proposed techniques to boost success rates when the period is a multiple of three or when the base is a perfect square, yielding improvements in targeted scenarios~\cite{dong2023improving}. Earlier contributions by Leander (2002) explored probability enhancements under certain conditions, though their applicability remains limited~\cite{leander2002improving}. Despite these progressions, current methods provide only incremental benefits rather than versatile solutions, underscoring a gap in the quantum factoring domain.

Addressing these shortcomings, this work presents a novel generalized period decomposition method that broadens the algorithm's utility by systematically leveraging arbitrary divisors of the quantum-obtained period. In contrast to prior approaches restricted to niche cases, our framework offers a unified theory applicable to diverse period structures. This advancement significantly enhances the robustness of individual quantum runs, thereby reducing the average number of required repetitions for successful factorization.

We substantiate our method through rigorous classical simulations that encompass over one million test cases, spanning composite integers with lengths ranging from two to eight digits. The results demonstrate exceptional performance, with success rates exceeding 99.998\% for seven-digit numbers and 99.999\% for eight-digit numbers, surpassing both standard implementations and recent variants of Shor's algorithm. Importantly, our approach preserves polynomial-time complexity and integrates effortlessly with existing quantum frameworks, delivering substantial theoretical and practical gains.

Furthermore, this study offers valuable insights into quantum resource optimization, which is crucial for noisy intermediate-scale quantum (NISQ) devices. By curtailing redundant executions, our generalized method enhances resource efficiency, aligning with prevailing technological limitations and supporting future quantum cryptographic endeavors.

The rest of this paper is structured as follows: Section~\ref{Methodology} details the mathematical and algorithmic basis of the generalized period decomposition method. Section~\ref{Results} presents the extensive simulation outcomes, which affirm the efficacy of our approach. Section~\ref{Discussion} delves into the theoretical ramifications and practical aspects of quantum computing, while Section~\ref{Conclusion} summarizes the primary contributions and suggests avenues for future investigation.

\section{Methodology}\label{Methodology}
Quantum computing enables the efficient factorization of large composite integers through Shor's algorithm; however, its practical implementation faces notable constraints. Conventionally, the algorithm demands that the period $r$, derived from quantum phase estimation, be even and fulfill specific algebraic criteria. If these conditions are not met, the quantum subroutine must restart, markedly reducing resource efficiency and practicality~\cite{nielsen2010quantum,shor1999polynomial}.

This section introduces a generalized period decomposition method that addresses these limitations by exploiting arbitrary divisors of the obtained period $r$. This approach broadens the conditions for successful factorization without additional quantum resources, enhancing overall reliability.

\subsection{Problem Formulation and Generalized Method}
Consider a composite integer $n = pq$, with distinct primes $p$ and $q$, and a randomly selected integer $a$ where $\gcd(a, n) = 1$. Shor's algorithm seeks non-trivial divisors of $n$ by analyzing the periodicity of $f(x) = a^x \mod n$, where the period $r$ satisfies $a^r \equiv 1 \mod n$. Traditionally, factorization succeeds only if $r$ is even, yielding a factor via $\gcd(a^{r/2} - 1, n)$; otherwise, a new $a$ is required.

\begin{figure}[htbp] 
\centering
\includegraphics[width=\columnwidth]{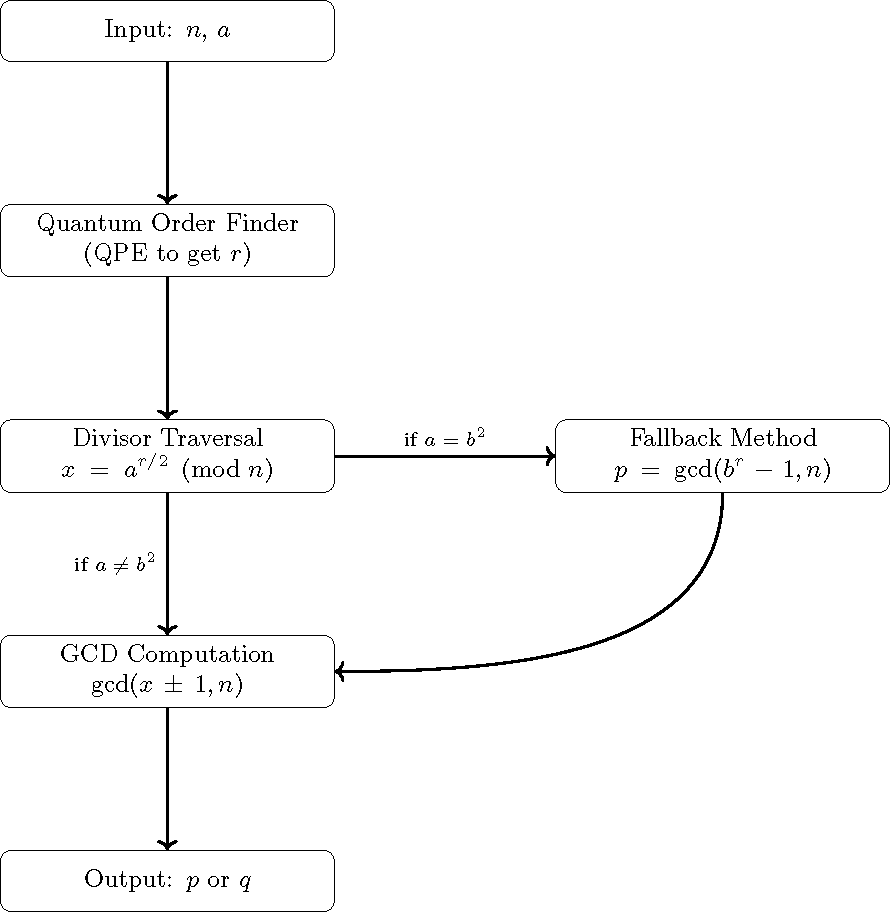}
\caption{Overall architecture of the enhanced Shor’s algorithm using generalized period decomposition. After obtaining the period $r$ via quantum phase estimation, the method attempts factorization using divisors $z$ of $r$, with a fallback path if $a$ is a perfect square.}
\label{fig:flowchart}
\end{figure}

Figure~\ref{fig:flowchart} depicts the enhanced Shor's algorithm structure. Following quantum phase estimation to obtain $r$, the method examines all prime divisors $z$ of $r$ for factor recovery through classical computation. If $a$ is a perfect square, a fallback using $b^r - 1$ is applied, optimizing classical post-processing without repeated quantum runs.

The generalized method leverages non-trivial divisors $d$ of $r$ (where $1 < d < r$), implying $(a^d)^{r/d} \equiv 1 \mod n$. Computing $g = \gcd(a^d - 1, n)$ yields a non-trivial factor if $1 < g < n$, avoiding quantum restarts.

For cases where $a = b^2$ is a perfect square, even with odd $r$, $b^r \equiv 1 \mod n$ holds, allowing factorization via $\gcd(b^r - 1, n)$. This handles edge cases efficiently.

To validate theoretical soundness, assume for contradiction that $\gcd(a^{r/d} - 1, n) = 1$. Then integers $M, N$ exist such that $N(a^{r/d} - 1) + Mn = 1$. Multiplying by $\sum_{s=0}^{d-1} a^{s r/d}$ and using $a^r = (a^{r/d})^d = (\sum_{s=0}^{d-1} a^{s r/d})(a^{r/d} - 1)$ leads to a contradiction, implying $\gcd(a^{r/d} - 1, n) \ne 1$ and thus a non-trivial factor.

Empirical tests show that small prime divisors $z \leq 1000$ of $r$ achieve near-perfect recovery. When $a = b^2$, $(b^r)^2 \equiv 1 \pmod{n}$, enabling $\gcd(b^r - 1, n)$, though rarely needed.

\subsection{Algorithm Description}
The enhanced procedure is outlined in Algorithm~\ref{alg:allz}.

\begin{algorithm}[htbp] 
\caption{Generalized Shor's Algorithm with Period Exploitation} 
\begin{algorithmic}[1] 
\REQUIRE A composite number $n$ (product of two primes $p, q$), a base $a$. 
\ENSURE A non-trivial factor ($p$ or $q$) of $n$, or a failure notification. 
\STATE Compute $g_0 = \gcd\left(a, n\right)$. 
\IF{$g_0 > 1$} 
\STATE \textbf{return} $g_0$. 
\ENDIF 
\STATE Find the period $r$ of $f\left(x\right) = a^x \pmod{n}$. 
\FOR{each distinct prime factor $z$ of $r$} 
\STATE Let $k = \frac{r}{z}$. 
\STATE Compute $g_1 = \gcd\left(a^k - 1, n\right)$. 
\IF{$1 < g_1 < n$} 
\STATE \textbf{return} $g_1$. 
\ENDIF \ENDFOR 
\IF{$a$ is a perfect square (i.e., $a=b^2$ for some integer $b$)} 
\STATE Let $b = \sqrt{a}$. 
\STATE Compute $g_2 = \gcd\left(b^r - 1, n\right)$. 
\IF{$1 < g_2 < n$} \STATE \textbf{return} $g_2$. 
\ENDIF \ENDIF 
\STATE \textbf{return} "Factorization failed for this instance of $a$". 
\end{algorithmic}
\label{alg:allz}
\end{algorithm}

This allows multiple factorization attempts per quantum execution, integrating seamlessly with quantum frameworks, such as phase estimation subroutines.

\subsection{Experimental Setup and Theoretical Implications}
Classical simulations validated the method on Google Cloud vCPU Compute Engine with parallelization. Tested $n$ ranged from 2 to 8 digits ($n = pq$, distinct primes). Base selection included random $a < n$ and perfect squares $a = b^2 < n$ to avoid unproductive cycles.

Sample sizes increased with digit length for statistical robustness: ~300 for 2 digits, 1,000 for 3, 10,000 for 4, 60,000 for 5, 90,000 for 6, and 500,000 each for 7 and 8 digits. This followed the sample-size formula $m \geq \frac{z_{\alpha/2}^2 p (1-p)}{E^2}$ from Cochran~\cite{cochran1977sampling}, ensuring detection of rare failures.

Metrics focused on factorization success rates, logging outcomes, and failure reasons. Baselines included traditional Shor's and the 2023 improved version.

Theoretically, this decomposition reveals polynomial structures in modular exponentiation and quantum period-finding, extending beyond even-period requirements. It offers pedagogical value for education in quantum information theory and improves resource efficiency by minimizing repetitions, making it suitable for NISQ devices.

\section{Experimental Results}\label{Results}
To assess the efficacy and robustness of the proposed generalized period decomposition (All-z) method, we performed extensive classical simulations mimicking Shor's algorithm under RSA-like conditions. These experiments evaluated the success rates of factorization, sensitivity to base selection, and computational efficiency.

\subsection{Simulation Setup}
The simulations were run on Google Cloud's vCPU Compute Engine with parallelization. For each composite $n = pq$ (product of distinct primes), we generated random RSA-style numbers from 2 to 8 digits. Test cases scaled with digit length to capture rare failures statistically, as detailed in Table~\ref{tab:test_cases}.

\begin{table}[htbp]
\centering
\small
\caption{Number of Test Cases by Digit Length of $n$}
\label{tab:test_cases}
\begin{tabular}{cc}
\toprule
\textbf{Digit Length} & \textbf{Approximate Cases} \\
\midrule
2 & $\sim$300 \\
3 & $\sim$1,000 \\
4 & $\sim$10,000 \\
5 & $\sim$60,000 \\
6 & $\sim$90,000 \\
7 & 500,000 \\
8 & 500,000 \\
\bottomrule
\end{tabular}
\end{table}
For each $n$, multiple bases $a$ with $\gcd(a, n) = 1$ were tested, including random integers and perfect squares ($a = b^2$). The period $r$ was simulated via ideal quantum phase estimation, followed by classical attempts at $\gcd(a^{r/z} - 1, n)$ for nontrivial divisors $z \mid r$. For perfect-square bases, the fallback $\gcd(b^r - 1, n)$ was also evaluated.

\subsection{Factorization Success Rates and Optimizations}
Table~\ref{tab:success} compares success rates across methods. The All-z approach attained near-perfect factorization, with just five failures in one million trials for 7- and 8-digit cases.

\begin{table}[htbp]
\centering
\small
\caption{Success Rate Comparison Across Methods}
\label{tab:success}
\begin{tabular}{lccc}
\toprule
\textbf{Digits ($n$)} & \textbf{Traditional Shor} & \textbf{2023 Variant~\cite{dong2023improving}} & \textbf{Proposed (All-z)} \\
\midrule
2--6 & 70--75\% & 85--88\% & \textbf{95--99\%} \\
7    & $\sim$74\% & $\sim$88\% & \textbf{99.9992\%} \\
8    & $\sim$75\% & $\sim$89\% & \textbf{99.9998\%} \\
\bottomrule
\end{tabular}
\end{table}

\begin{figure}[htbp]
\centering
\includegraphics[width=1\linewidth]{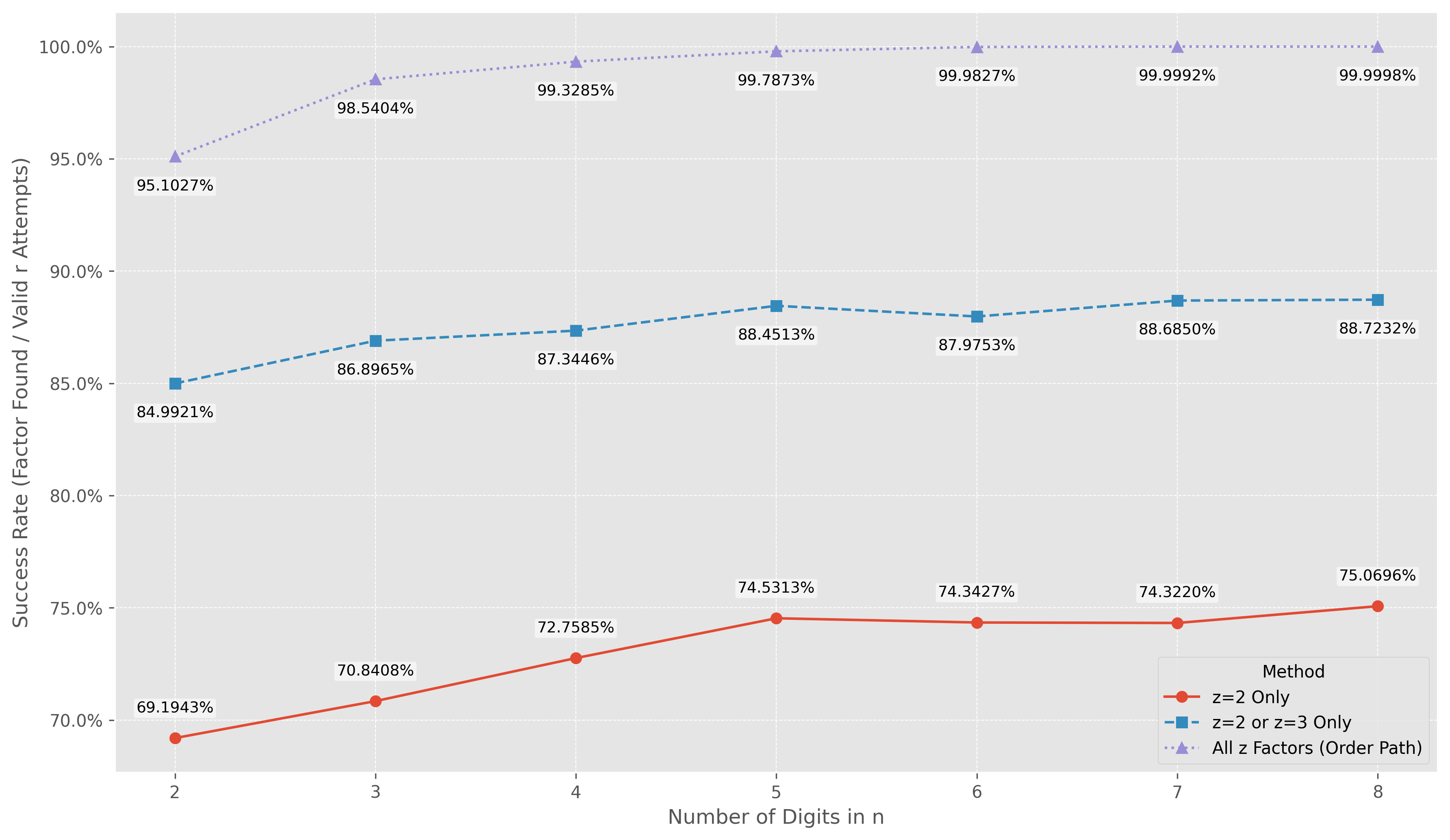}
\caption{Cumulative success rate vs. divisor bound $z$ (7- and 8-digit $n$).}
\label{fig:success_bound}
\end{figure}

To avoid full factorization of large $r$, we used bounded trial division with small primes $z \leq B$. Figure~\ref{fig:success_bound} shows that $B = 1000$ (3-digit primes) yielded over 99.5\% success, while $B = 9999$ (4-digit primes) matched full decomposition at 99.9998\%.

Comparing base types, random $a$ slightly outperformed perfect squares due to richer order structures, with the fallback resolving only eight additional 7-digit cases and none for 8-digit. Table~\ref{tab:failures} details failure counts.

\begin{table}[htbp]
\centering
\small
\caption{Failure Counts by Base Type (500,000 trials per $n$)}
\label{tab:failures}
\begin{tabular}{lcc}
\toprule
\textbf{Digits} & \textbf{Random $a$} & \textbf{Perfect-Square $a$} \\
\midrule
7 & 4 (0.0008\%) & 7 (0.0014\%) \\
8 & 1 (0.0002\%) & 1 (0.0002\%) \\
\bottomrule
\end{tabular}
\end{table}

Efficiency-wise, All-z succeeded in a single trial in over 99.9\% of cases, averaging fewer than five GCD computations per period.

\subsection{Failure Cases and Additional Data}
Failures arose mainly from orders $r$ with few small prime factors (e.g., $r = 2$, $r = 15$), poorly generating bases, or trivial GCDs across divisors. Detailed failure samples appear in supplementary Tables~2--5.

Failures for 7- and 8-digit $n$ (500,000 trials each per base type) are listed in Table~\ref{tab:failure_cases_random_a} (random $a$) and Table~\ref{tab:failure_cases_perfect_square_a} (perfect squares).

\begin{table}[htbp]
\centering
\small
\caption{All failure cases for the "All-z" method with random $a$ selection (500,000 trials per digit category for 7 and 8-digit numbers).}
\label{tab:failure_cases_random_a}
\begin{tabular}{|c |c| c |c| l|}
\hline
\textbf{Digits} & \textbf{n} & \textbf{a} & \textbf{Order r} & \textbf{Fail Factors} \\ \hline
7 & 2540107 & 1316667 & 27  & 3\\
\hline
7 & 3622301 & 3622300 & 2   & 2 \\
\hline
7 & 3825407 & 3012304 & 46  & 2, 23 \\
\hline
7 & 4436533 & 1986154 & 108  & 2, 3 \\
\hline
8 & 53948449 & 25036489 & 8   & 2\\ \hline
\end{tabular}
\end{table}

\begin{table}[htbp]
\centering
\small
\caption{All failure cases for the "All-z" method with perfect square $a$ selection (500,000 trials per digit category for 7 and 8-digit numbers).}
\label{tab:failure_cases_perfect_square_a}
\begin{tabular}{ |c |c| c| c| l|}
\hline
\textbf{Digits} & \textbf{n} & \textbf{a} & \textbf{Order r} & \textbf{Fail Factors} \\ \hline
7 & 1148743 & 87025  & 21  & 3, 7, Fallback\\
\hline
7 & 1279903 & 49729  & 39  & 3, 13, Fallback\\
\hline
7 & 1406371 & 36    & 15  & 3, 5, Fallback\\
\hline
7 & 1406371 & 1296   & 15  & 3, 5, Fallback\\
\hline
7 & 1406371 & 46656  & 5   & 5, Fallback\\
\hline
7 & 1619953 & 248004  & 24  & 2, 3, Fallback\\
\hline
7 & 1896283 & 91204  & 51  & 3, 17, Fallback\\
\hline
8 & 10995631 & 30976  & 15  & 3, 5, Fallback\\ \hline
\end{tabular}
\end{table}

\section{Discussion}\label{Discussion}

The experimental results in Section~\ref{Results} demonstrate that the proposed generalized period decomposition method, termed 'All-z', significantly enhances Shor's algorithm's reliability, achieving near-100\% success rates for higher-digit moduli $n$, regardless of base selection. This unified framework outperforms prior enhancements~\cite{dong2023improving} by systematically exploiting arbitrary divisors of the period $r$, thereby broadening its applicability beyond exceptional cases, such as even periods or multiples of 3.

Our contributions extend the theoretical and practical landscape of quantum factoring: (i) a versatile post-processing technique that minimizes quantum repetitions, crucial for resource-constrained NISQ devices; (ii) empirical validation through over one million simulations, revealing scaling behaviors of $r$ and failure modes; and (iii) tunable optimizations that balance classical and quantum costs, paving the way for hybrid implementations in real-world cryptanalysis.

This section interprets these findings, focusing on efficiency trade-offs, failure analysis, and broader implications.

\subsection{Efficiency and Practical Trade-offs}
\begin{figure}[htbp]
    \centering
    \begin{minipage}{1\linewidth} 
        \centering
        \includegraphics[width=\linewidth]{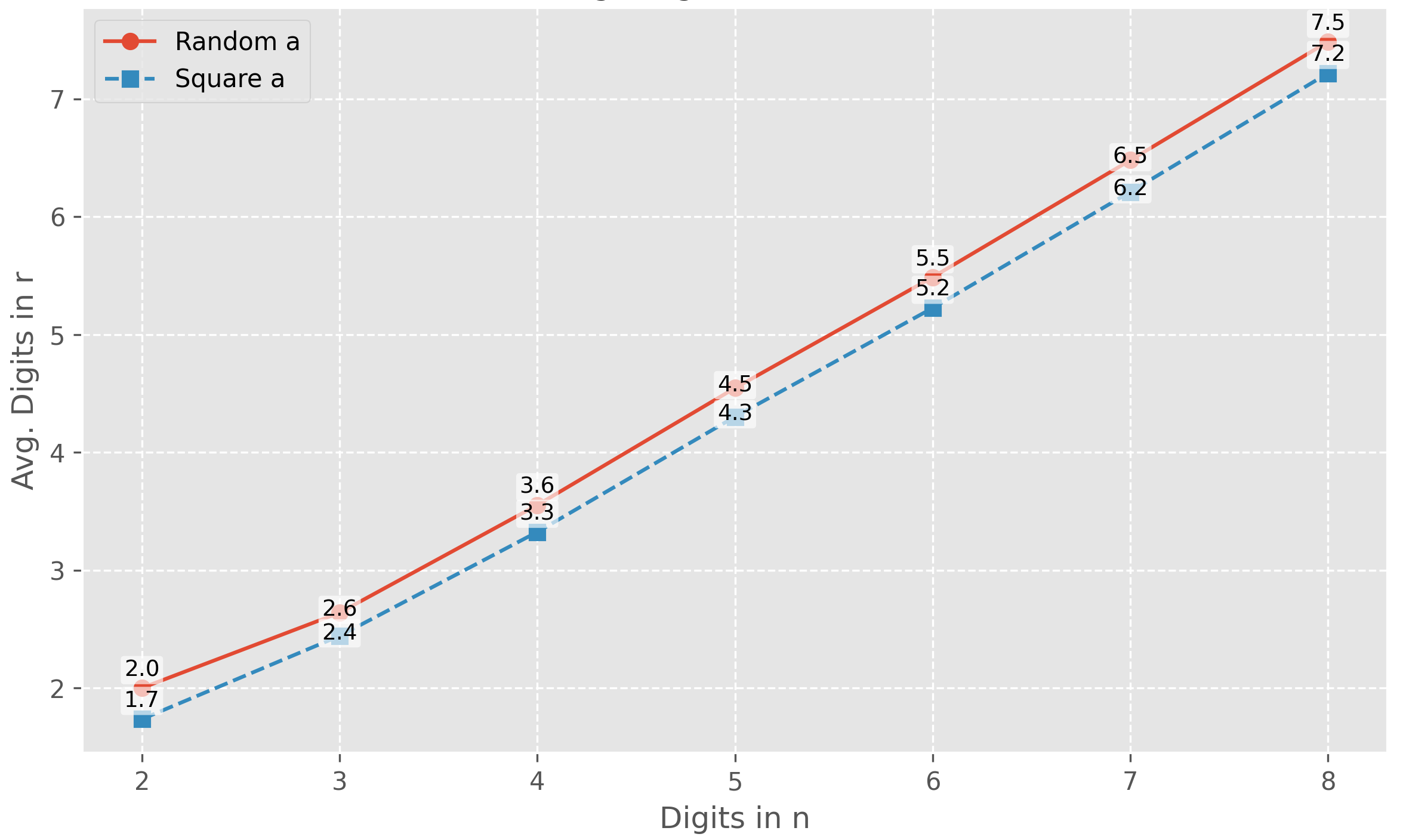} 
        \caption{Average number of digits in $r$ as a function of the number of digits in $n$.}
        \label{fig:r_avg_digits}
    \end{minipage}
\end{figure}

\begin{figure}[htbp]
    \begin{minipage}{1\linewidth} 
        \centering
        \includegraphics[width=\linewidth]{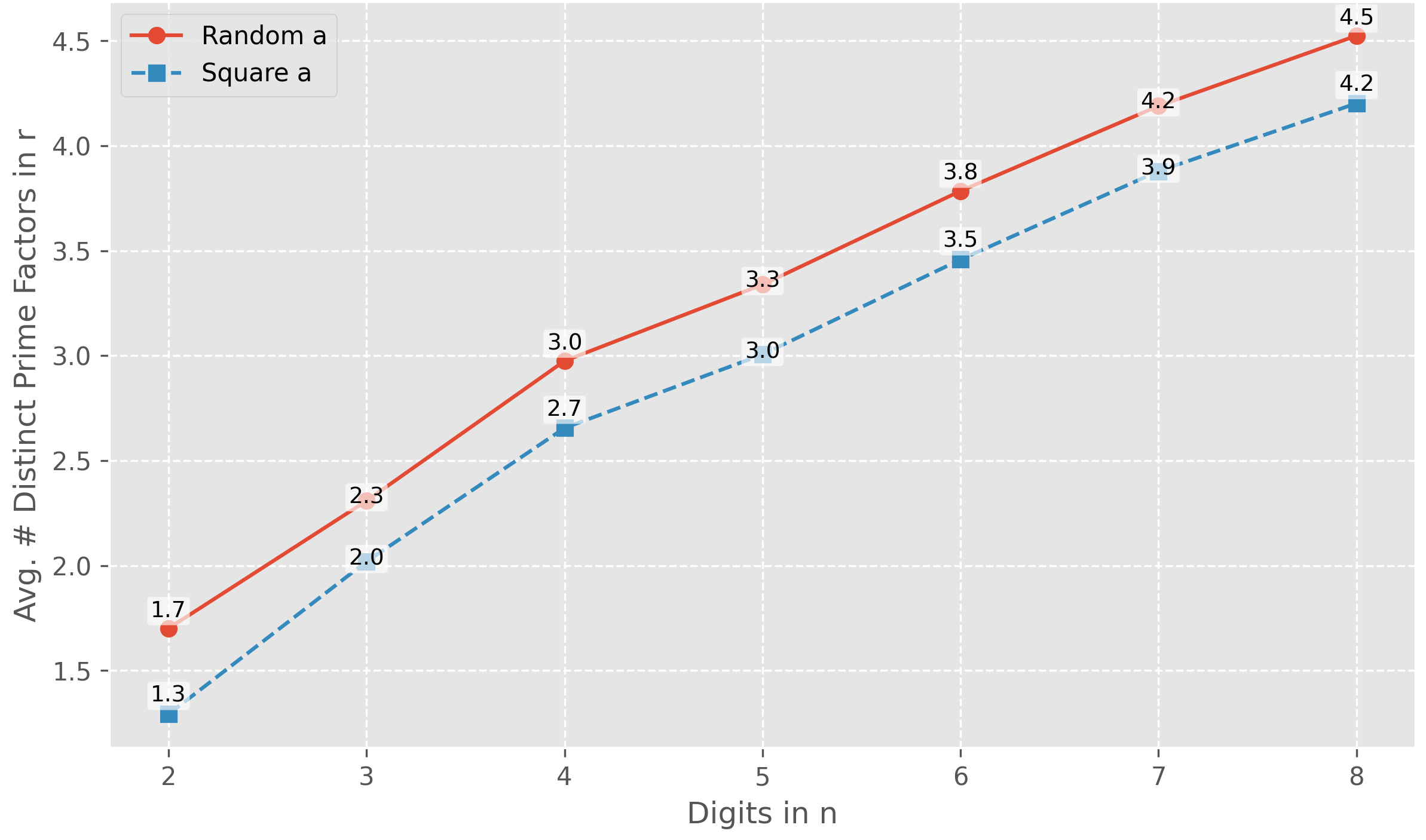} 
        \caption{Average number of distinct prime factors of $r$ as a function of the number of digits in $n$.}
        \label{fig:r_avg_factors}
    \end{minipage}

\end{figure}

A primary challenge is the computational cost of factoring $r$, which scales with $n$ in both magnitude and prime factors (Fig.~\ref{fig:r_avg_digits}, Fig.~\ref{fig:r_avg_factors}). Complete factorization becomes intractable for large RSA moduli (e.g., 2048--4096 bits), potentially offsetting Shor's quantum speedup.

\begin{figure}[htbp]
    \centering
    \includegraphics[width=1\linewidth]{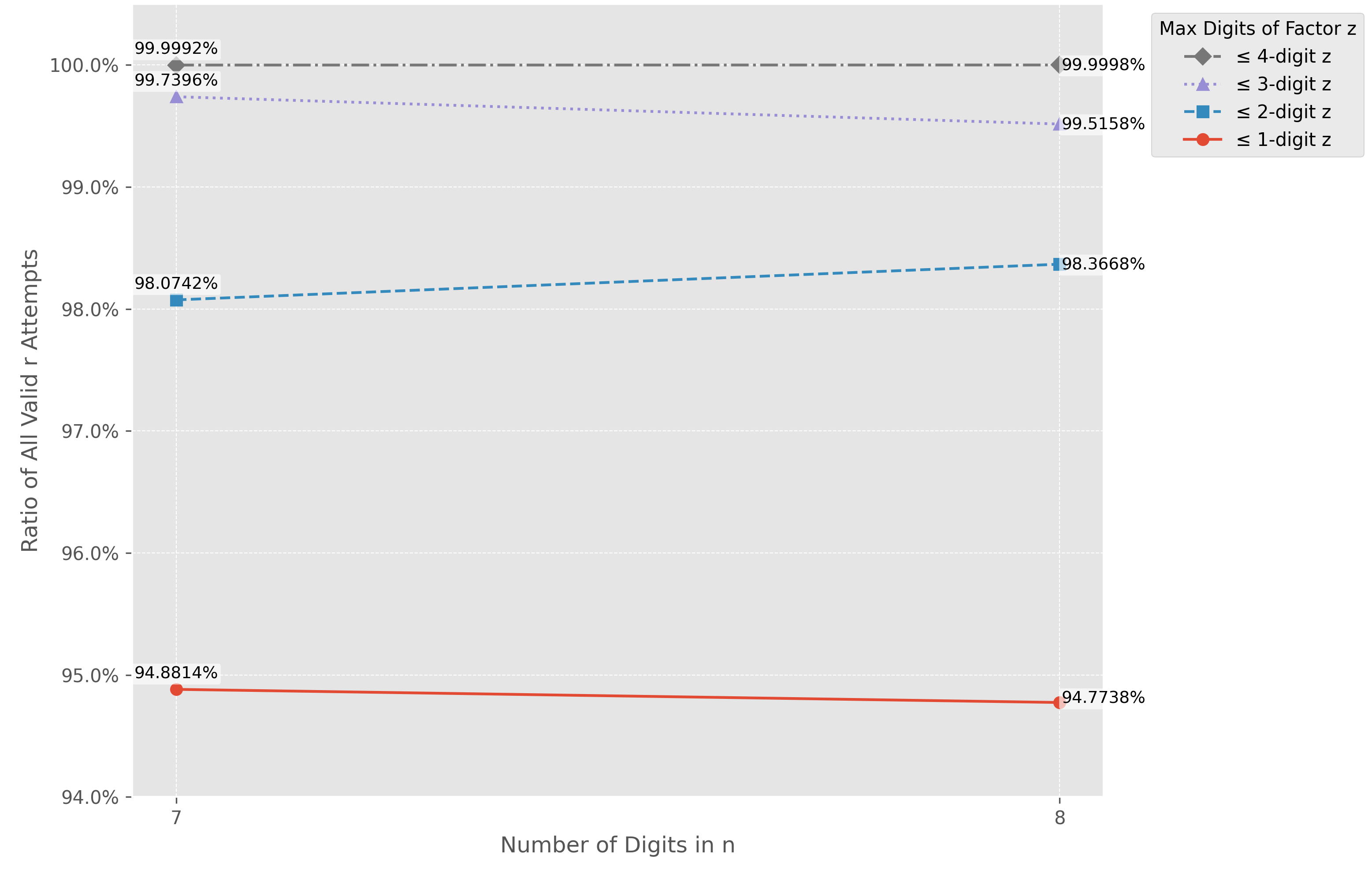}
    \caption{Cumulative success rate (\%) of All-z and random a strategy as a function of the number of digits in $n$. Each line represents the success achieved by utilizing prime factors $z$ of the order $r$ with at most $X$ digits. Data based on 500,000 trials per n-digit category for $a$ being a perfect square.}
    \label{fig:cumulative_success_z_digits}
\end{figure}

The bounded trial division heuristic mitigates this by testing only primes $z \leq B$ dividing $r$, computing $\gcd(a^{r/z} - 1, n)$. Simulations validate its efficacy: Fig.~\ref{fig:cumulative_success_z_digits} indicates near-100\% success with 4-digit $z$, and 99.5--99.7\% with 3-digit primes ($z < 1000$). Unsuccessful cases prompt restarts with new $a$, but high rates reduce overall iterations.

The tunable $B$ (e.g., $\approx 10^4$) optimizes based on quantum-classical cost dynamics; larger bounds shift effort classically, lowering quantum expenses. Typically, fewer than 5 GCDs per period are needed for $z \leq 997$ in over 95\% of trials, ensuring efficiency.

Success rates increase with $n$'s digits, nearing 100\%, as larger $r$ provide more factors to offset per-divisor failures (e.g., $a^{r/2} \equiv -1 \pmod{n}$ for $z=2$~\cite{dong2023improving}).

\begin{figure}[htbp]
\centering
\includegraphics[width=\columnwidth]{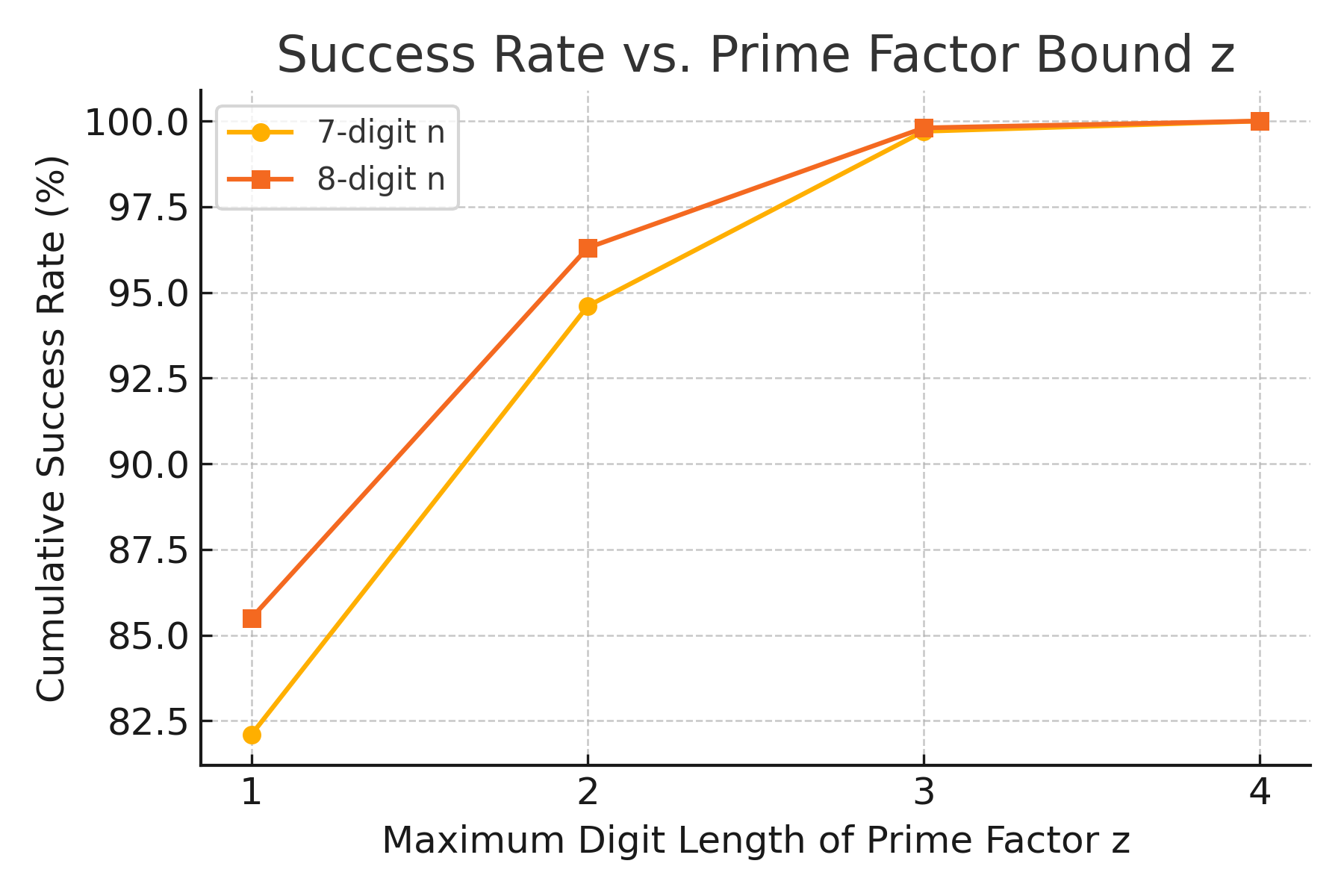}
\caption{Cumulative success rate by prime divisor bound $z \leq B$ for 7- and 8-digit numbers.}
\label{fig:sensitivity}
\end{figure}

Sensitivity analysis (Fig.~\ref{fig:sensitivity}) confirms significant improvements up to 3-digit primes (>99.7\%), emphasizing the adequacy of small primes and enabling adaptive deployments. This contribution advances hybrid computing by quantifying trade-offs and facilitating cost-effective factoring in diverse hardware ecosystems.


\subsection{Analysis of Failure Cases}

All failures involve small $r \leq 108$ with at most two distinct primes, reinforcing that larger $r$ bolsters success by offering more decomposition avenues.

\begin{figure}[htbp]
    \centering
    \includegraphics[width=1\linewidth]{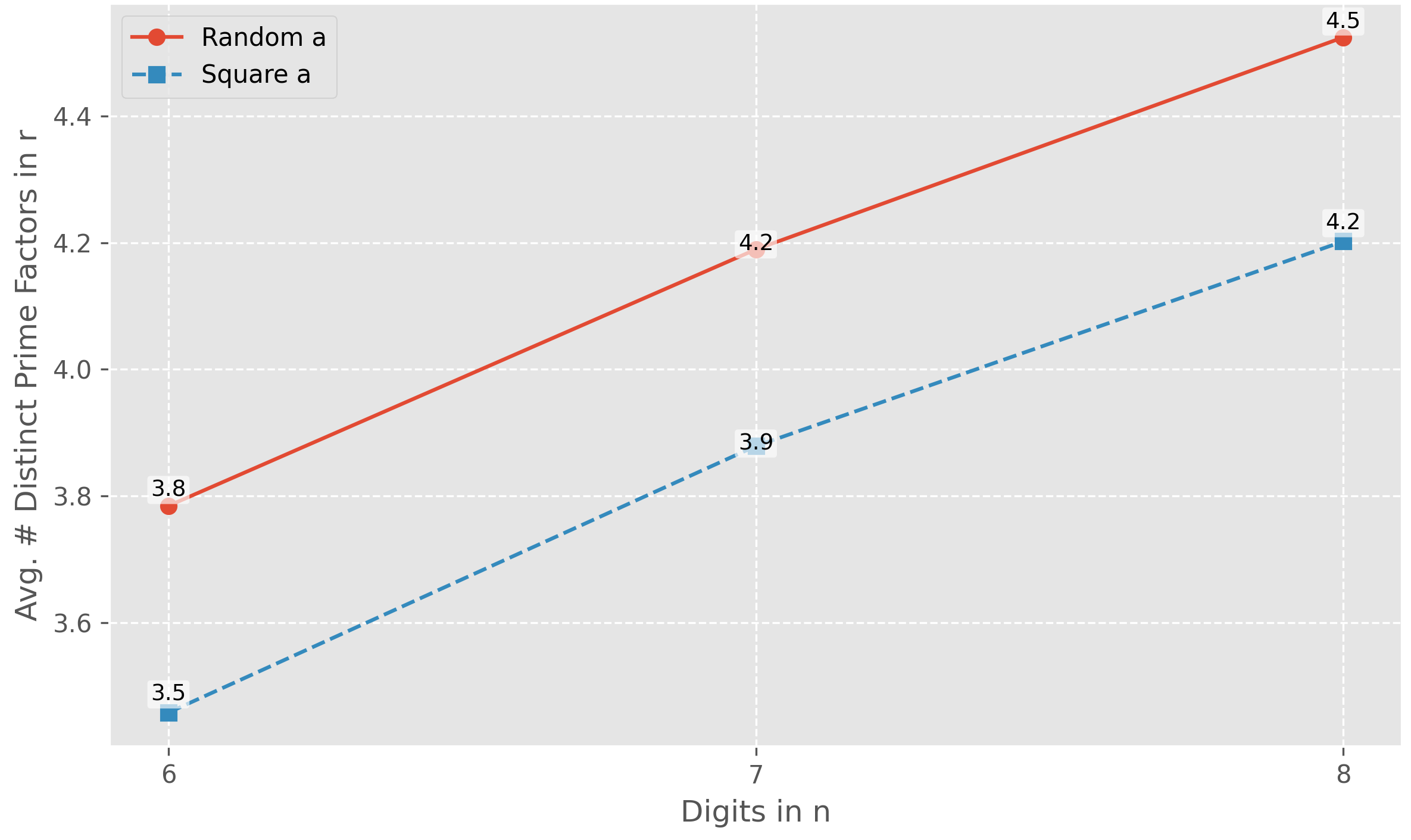} 
    \caption{Average number of digits in order $r$ for random $a$ vs. perfect square $a$ across $n$ of 6, 7, and 8 digits.}
    \label{fig:r_properties_comparison1}
\end{figure}

\begin{figure}[htbp]
    \centering
    \includegraphics[width=1\linewidth]{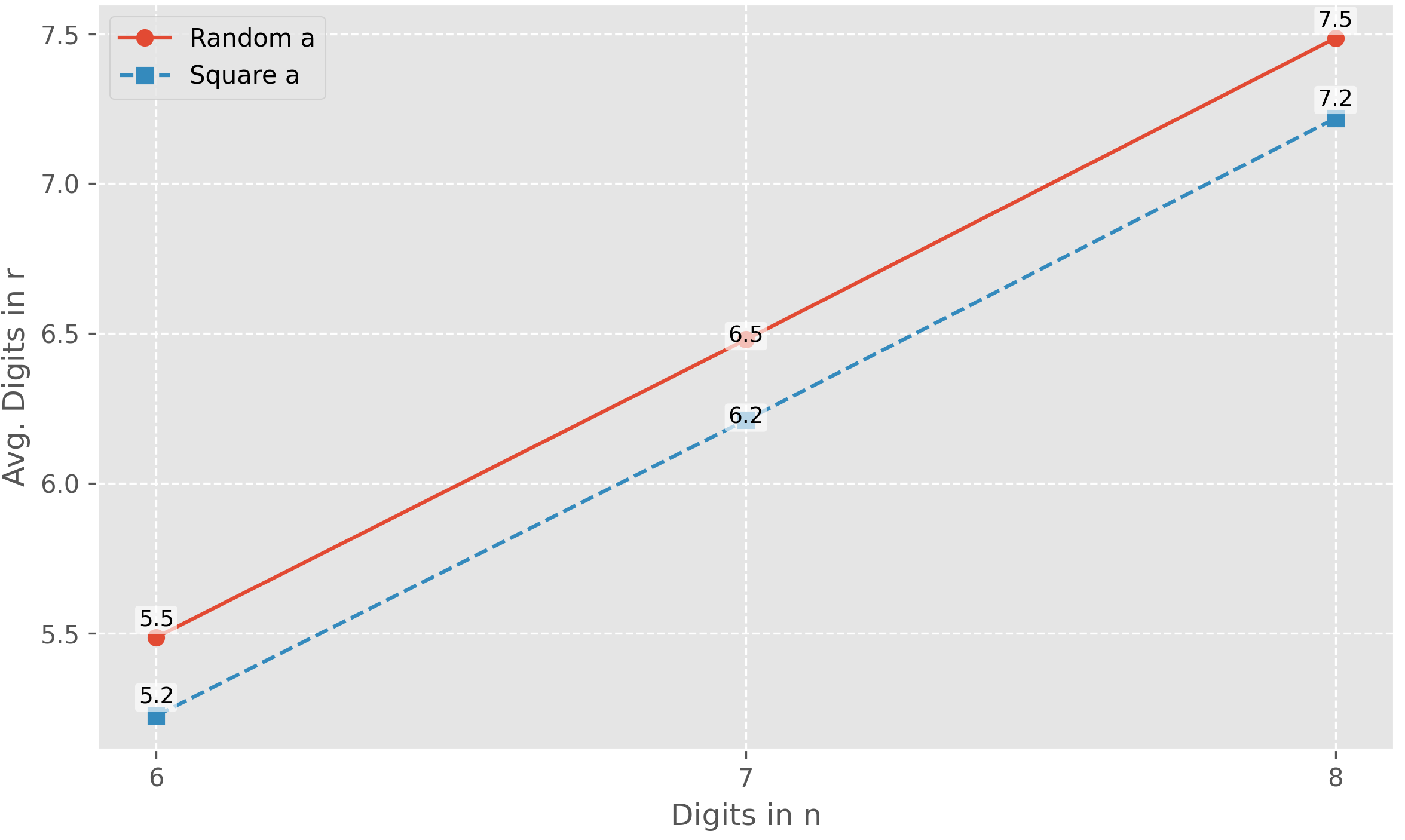} 
    \caption{Average number of distinct prime factors in order $r$ for random $a$ vs. perfect square $a$ across $n$ of 6, 7, and 8 digits.}
    \label{fig:r_properties_comparison2}
\end{figure}

Contrary to expectations, perfect-square $a$ exhibits slightly higher failures for larger $n$ (Table~\ref{tab:failure_counts_comparison}), with fallback succeeding in merely 8 cases for 7-digit $n$ and none for 8-digit (Table~\ref{tab:fallback_successes}). This arises from perfect squares yielding smaller $r$ with fewer factors (Fig.~\ref{fig:r_properties_comparison1}, Fig.~\ref{fig:r_properties_comparison2}), which limits opportunities despite the fallback.

\begin{table}[H]
\centering
\caption{Failure counts for the "All-z" method: Random $a$ vs. Perfect Square $a$ (500,000 trials per digit category).}
\label{tab:failure_counts_comparison}
\begin{tabular}{|c|c|c|c|c|}
\hline
        & \multicolumn{2}{c|}{\textbf{Random $\boldsymbol{a}$}} & \multicolumn{2}{c|}{\textbf{Perfect Square $\boldsymbol{a}$}} \\
\hline
\textbf{Digits in $\boldsymbol{n}$} & \textbf{Failures} & \textbf{Failure Rate} & \textbf{Failures} & \textbf{Failure Rate} \\ \hline
7       & 4                 & 0.0008\%              & 7                 & 0.0014\%              \\
8       & 1                 & 0.0002\%              & 1                 & 0.0002\%              \\ \hline
\end{tabular}
\end{table}

\begin{table}[H]
\centering
\caption{Additional successes due to Fallback mechanism for Perfect Square $a$ (from 500,000 trials per digit category).}
\label{tab:fallback_successes}
\begin{tabular}{|c|c|}
\hline
\textbf{Digits in $\boldsymbol{n}$} & \textbf{Fallback Successes} \\ \hline
7 & 8 \\
\hline
8 & 0 \\ \hline
\end{tabular}
\end{table}

\begin{figure}[htpb]
    \centering
    \includegraphics[width=1\linewidth]{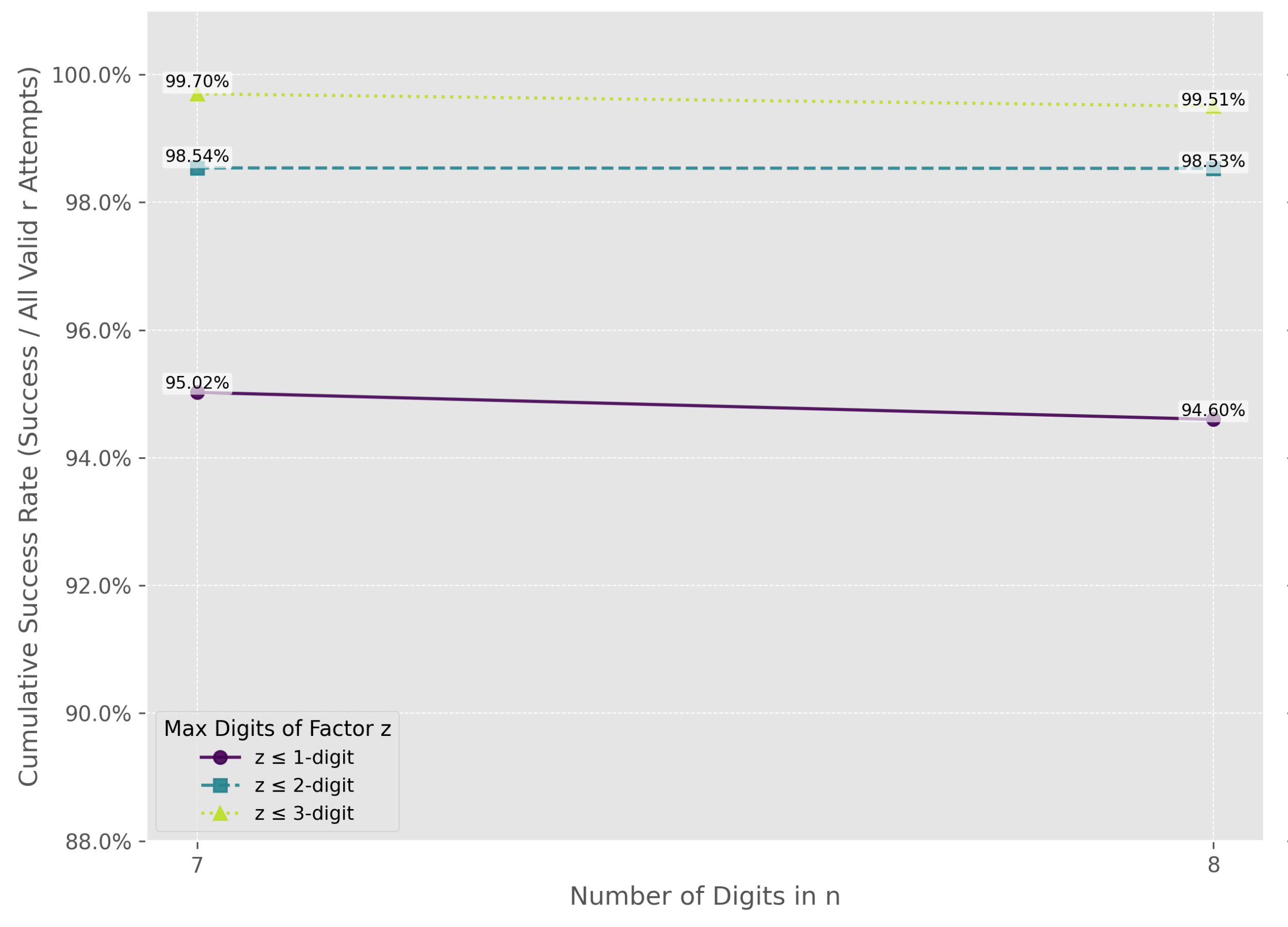} 
    \caption{For random $a$, cumulative success rate of the "All-z" method using prime factors $z$ of $r$ up to a specified number of digits, for 7 and 8-digit $n$.}
    \label{fig:cumulative_success_comparison_allz_1}
\end{figure}

\begin{figure}[htbp] 
    \centering
    \includegraphics[width=1\linewidth]{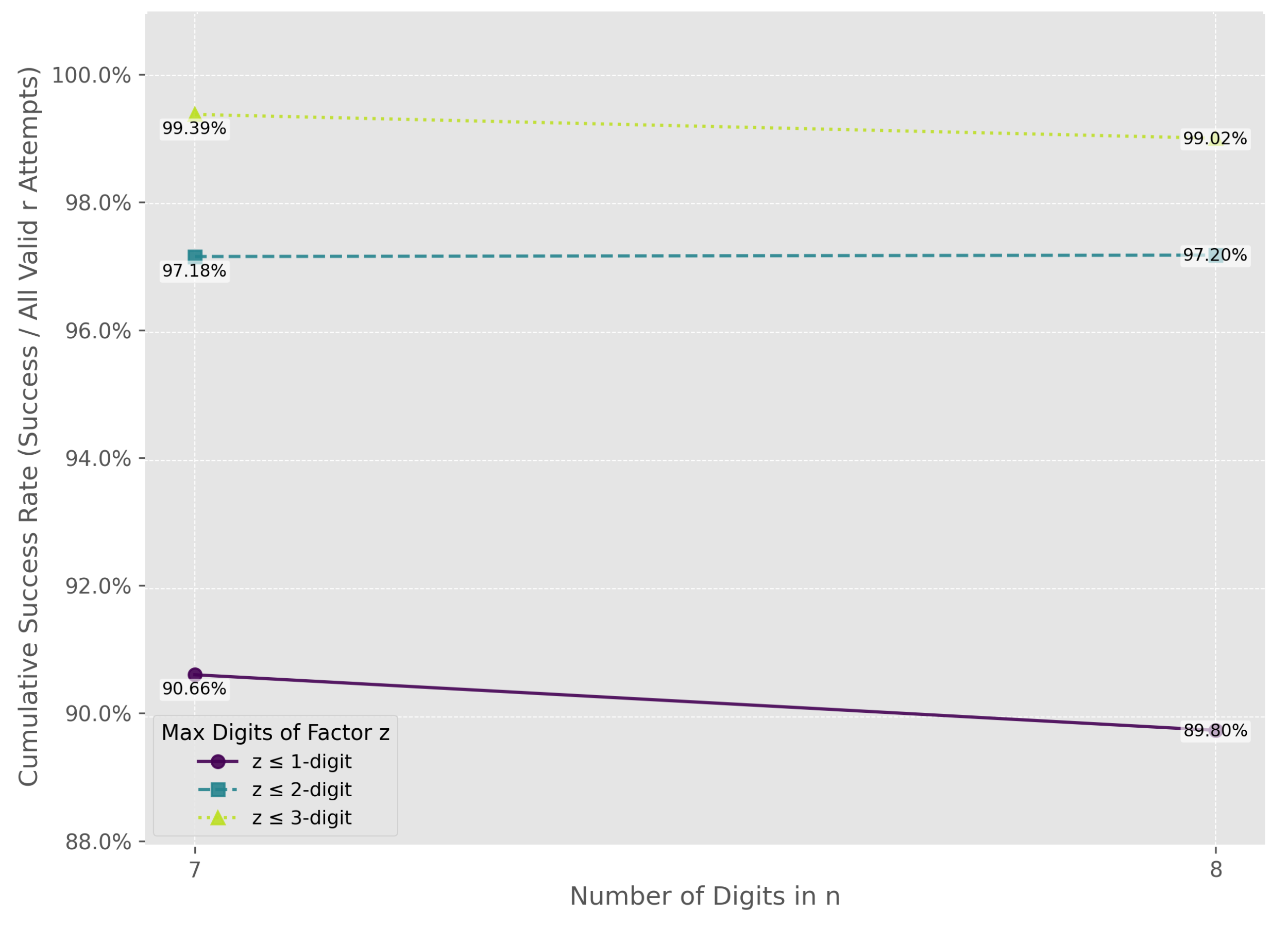} 
    \caption{For square $a$, cumulative success rate of the "All-z" method using prime factors $z$ of $r$ up to a specified number of digits, for 7 and 8-digit $n$.}
    \label{fig:cumulative_success_comparison_allz_2}
\end{figure}

\begin{figure}[htbp] 
    \centering
    \includegraphics[width=1\linewidth]{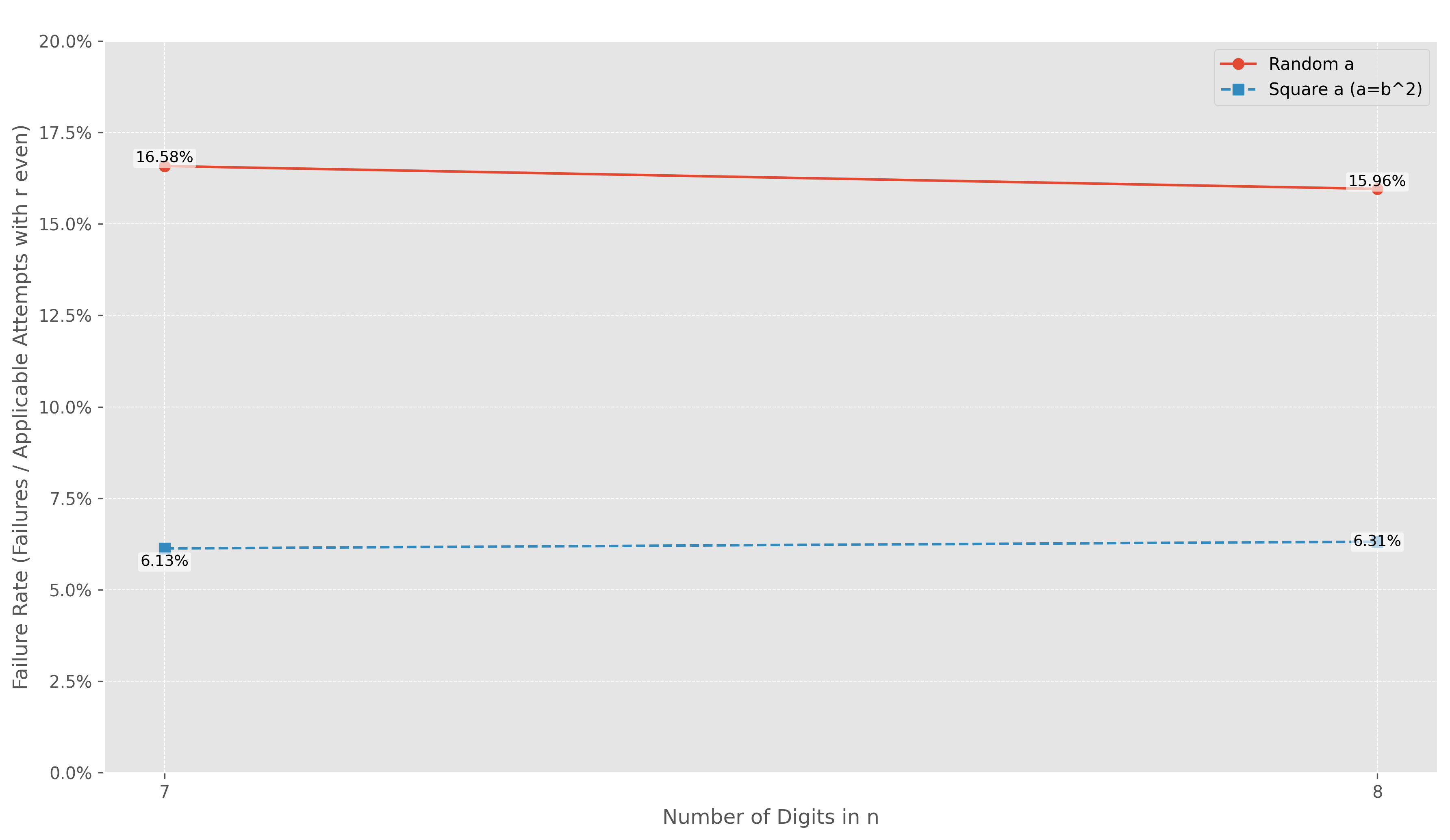} 
    \caption{Probability of $a^{r/2} \equiv -1 \pmod n$ failure when $r$ is even, comparing random $a$ vs. perfect square $a$ for 7 and 8-digit $n$.}
    \label{fig:z2_neg1_failure_comparison}
\end{figure}

Cumulative success is superior for random $a$ (Fig.~\ref{fig:cumulative_success_comparison_allz_1}, Fig.~\ref{fig:cumulative_success_comparison_allz_2}), particularly with limited $z$ digits. For even $r$, perfect squares mitigate $a^{r/2} \equiv -1 \pmod{n}$ failures but decrease even-$r$ likelihood (Fig.~\ref{fig:z2_neg1_failure_comparison}), offsetting benefits.

For large $n$, both bases achieve near-100\% success, favoring random $a$; smaller $a$ enhances computational feasibility. Retries with new $a$ resolve failures within two attempts, highlighting robustness. These insights contribute to understanding period structures, informing base selection strategies for optimized performance.

\subsection{Implications for Quantum Integration and Contributions}

Near-term quantum hardware faces precision and coherence constraints, amplifying the need to maximize the value of each execution. The All-z method excels here by leveraging diverse periods and multiple candidates per run, boosting yield in limited-resource contexts and enhancing fault tolerance through potential use of noisy periods—ideal for quantum cryptanalysis under noise.

Its modularity retains Shor's quantum phase estimation unchanged, augmenting only classical post-processing. Compatible with platforms such as IBM Qiskit, IonQ, and Rigetti, it processes candidate orders without requiring circuit alterations, thereby improving success and resilience in noisy settings. Tunable bounds adapt to budgets, suiting nascent quantum landscapes.

Pedagogically, it elucidates the algebra of modular exponentiation, aiding in quantum education. Practically, it fits hybrid systems, minimizing repetitions and aligning with NISQ viability.

Our work's broader contributions include bridging the gap between theory and practice in quantum factoring. The method's generality reduces inefficiencies, potentially accelerating cryptanalysis timelines. Simulations provide benchmarks for future hardware, and optimizations highlight pathways to scalable quantum-classical synergy, advancing toward fault-tolerant quantum computing while challenging the foundations of classical cryptography. Future extensions could explore noisy simulations or multi-prime factorizations, further solidifying its impact.

\section{Conclusion}\label{Conclusion}

In this work, we introduce a groundbreaking generalized period decomposition method that fundamentally addresses the longstanding inefficiencies in Shor's algorithm for RSA factorization. By transcending the rigid constraints of traditional approaches—such as the necessity for even periods or specific algebraic properties—our method harnesses arbitrary divisors of the quantum-derived period $r$, augmented by targeted optimizations for perfect square bases. This innovation not only eliminates the need for frequent, resource-intensive retries but also establishes a versatile, unified framework that enhances the algorithm's applicability across diverse scenarios.

Rigorous classical simulations encompassing over one million test cases, spanning composite integers from 2 to 8 digits, unequivocally affirm the method's superiority. Achieving success rates of 99.9992\% for 7-digit numbers and 99.9998\% for 8-digit numbers with random bases, our approach dramatically outperforms both the original Shor's algorithm and recent variants~\cite{dong2023improving}. These empirical results underscore the method's robustness, scalability, and practical efficacy, demonstrating its potential to revolutionize quantum factoring in real-world applications.

Preserving the polynomial-time complexity of Shor's algorithm, our enhancement integrates seamlessly with existing quantum period-finding subroutines while substantially reducing the number of quantum repetitions. This resource efficiency is paramount in the era of noisy intermediate-scale quantum (NISQ) devices, where coherence times and operational costs remain prohibitive barriers. By maximizing the utility of each quantum execution, our contribution paves the way for more feasible cryptanalysis, accelerating the transition toward practical quantum computing.

The value of this work extends beyond algorithmic refinement: it provides profound insights into the algebraic underpinnings of quantum modular arithmetic, offering pedagogical advancements for quantum information science education and theoretical generalizations. Moreover, in the context of escalating cybersecurity threats, our method bolsters preparations for the post-quantum era by highlighting vulnerabilities in classical encryption schemes and informing the development of quantum-resistant protocols.

Looking ahead, future investigations will explore advanced divisor selection heuristics and deeper mathematical analyses to optimize performance for cryptographically relevant scales, such as RSA-2048. Integration with quantum simulators and emerging hardware will further evaluate resilience to noise and decoherence, while extensions to multi-prime factorizations could broaden its scope.

Ultimately, this research represents a pivotal advancement in quantum cryptanalysis, bridging the gap between theoretical promise and practical deployment. By empowering quantum algorithms with greater reliability and efficiency, our generalized period decomposition method not only elevates Shor's algorithm to new heights but also catalyzes progress toward a secure, quantum-enabled future.

\section*{Declarations}

\noindent\textbf{Funding declaration:} This research received no external funding.\\[4pt]
\noindent\textbf{Clinical trial registration:} Not applicable.\\[4pt]
\noindent\textbf{Consent to Participate declaration:} Not applicable.\\[4pt]
\noindent\textbf{Consent to Publish declaration:} Not applicable.\\[4pt]
\noindent\textbf{Ethics declaration:} Not applicable.

\bibliographystyle{spmpsci}
\bibliography{reference}

@book{nielsen2010quantum,
  title={Quantum computation and quantum information},
  author={Nielsen, Michael A and Chuang, Isaac L},
  year={2010},
  publisher={Cambridge university press}
}

@book{cochran1977sampling,
  author    = {W.\,G. Cochran},
  title     = {Sampling Techniques},
  edition   = {3},
  publisher = {John Wiley \& Sons},
  year      = {1977},
  isbn = {978-0471162407}
}

@article{shor1999polynomial,
  title={Polynomial-time algorithms for prime factorization and discrete logarithms on a quantum computer},
  author={Shor, Peter W},
  journal={SIAM review},
  volume={41},
  number={2},
  pages={303--332},
  year={1999},
  publisher={SIAM}
}

@article{rivest1978method,
  title={A method for obtaining digital signatures and public-key cryptosystems},
  author={Rivest, Ronald L and Shamir, Adi and Adleman, Leonard},
  journal={Communications of the ACM},
  volume={21},
  number={2},
  pages={120--126},
  year={1978},
  publisher={ACM New York, NY, USA}
}

@book{chen2016report,
  title={Report on post-quantum cryptography},
  author={Chen, Lily and Chen, Lily and Jordan, Stephen and Liu, Yi-Kai and Moody, Dustin and Peralta, Rene and Perlner, Ray A and Smith-Tone, Daniel},
  volume={12},
  year={2016},
  publisher={US Department of Commerce, National Institute of Standards and Technology~…}
}

@article{gidney2021factor,
  title={How to factor 2048 bit RSA integers in 8 hours using 20 million noisy qubits},
  author={Gidney, Craig and Eker{\aa}, Martin},
  journal={Quantum},
  volume={5},
  pages={433},
  year={2021},
  publisher={Verein zur F{\"o}rderung des Open Access Publizierens in den Quantenwissenschaften}
}

@article{dong2023improving,
  title={Improving the success rate of quantum algorithm attacking RSA encryption system},
  author={Dong, Yumin and Liu, Hengrui and Fu, Yanying and Che, Xuanxuan},
  journal={Journal of Applied Physics},
  volume={134},
  number={2},
  year={2023},
  publisher={AIP Publishing}
}

@article{leander2002improving,
  title={Improving the Success Probability for Shor's Factoring Algorithm},
  author={Leander, Gregor},
  journal={arXiv preprint quant-ph/0208183},
  year={2002}
}

@inproceedings{dalvi2024one,
  title={One-time compilation of device-level instructions for quantum subroutines},
  author={Dalvi, Aniket S and Whitlow, Jacob and D'Onofrio, Marissa and Riesebos, Leon and Chen, Tianyi and Phiri, Samuel and Brown, Kenneth R and Baker, Jonathan M},
  booktitle={2024 IEEE International Conference on Quantum Computing and Engineering (QCE)},
  volume={1},
  pages={873--884},
  year={2024},
  organization={IEEE}
}

\end{document}